\documentclass[twocolumn,prb,aps,superscriptaddress,floatfix]{revtex4}
\usepackage{graphicx}
\usepackage{amsmath,amssymb,amsfonts}

\begin{document}  
\title{Optical properties of charged quantum dots doped with a single
  magnetic impurity} 

\author{U. C. Mendes} 
\affiliation{Quantum Theory Group, Security and Disruptive Technologies, National Research Council, Ottawa, Canada K1A0R6}  
\affiliation{Institute of Physics ``Gleb Wataghin", State University of Campinas, Campinas-SP, Brazil}

\author{M. Korkusinski} 
\affiliation{Quantum Theory Group, Security and Disruptive Technologies, National Research Council, Ottawa, Canada K1A0R6}  

\author{A. H. Trojnar} 
\affiliation{Quantum Theory Group, Security and Disruptive Technologies, National Research Council, Ottawa, Canada K1A0R6} 
\affiliation{Department of Physics, University of Ottawa, Ottawa, Canada}
 
\author{P. Hawrylak}  
\affiliation{Quantum Theory Group, Security and Disruptive Technologies, National Research Council, Ottawa, Canada K1A0R6} 
\affiliation{Department of Physics, University of Ottawa, Ottawa, Canada}
     
\begin{abstract} 
We present a microscopic theory of the optical properties of
self-assembled quantum dots doped with a single magnetic manganese
(Mn) impurity and containing a controlled number of electrons. 
The single-particle electron and heavy-hole electronic shells are
described by two-dimensional harmonic oscillators. 
The electron-electron, electron-hole Coulomb  as well as the
short-range electron spin-Mn spin and hole spin-Mn spin contact
exchange interactions are included.  
The electronic states of the photo-excited electron-hole-Mn complex
and of the final electron-Mn complex are expanded in a finite number
of configurations and the full interacting Hamiltonian is diagonalized
numerically.  
The emission spectrum is predicted as a function of photon energy for
a given number of electrons and different number of confined
electronic quantum dot shells.  
We show how emission spectra allow to identify the number of
electronic shells, the number of electrons populating these shells
and, most importantly, their spin. We show that electrons not interacting directly 
with the spin of Mn ion do so via electron-electron interactions. This indirect 
interaction is a strong effect even when Mn impurity is away from the quantum dot 
center.   
\end{abstract}

\maketitle 


\section{Introduction}

There is currently interest in developing means of controlling spin 
at the nanoscale. \cite{hawrylak_grabowski_prb1991,loth_heinrich_natphys2010,ochsenbein_gamelin_natnano2009,bussian_crooker_natmat2009,dietl_natmat2010,gaj_kossut_book2011,koenraad_flatte_nat2011}
This includes spin of electrons and holes in
gated\cite{henneberger_benson_book2008,hsieh_shim_rpp2012} 
and self-assembled quantum 
dots\cite{michler-book2003,kioseoglou_jonker_prl2008} 
as well as magnetic impurities in semiconductors.\cite{dietl_natmat2010,gaj_kossut_book2011}
It is now possible to place, optically detect, 
and manipulate a single  magnetic impurity in a single self-assembled  
quantum dot
(QD).\cite{besombes_leger_prl2004,hundt_henneberger_prb2004,besombes_leger_prb2005,kudelski_lemaitre_prl2007,goryca_kazimierczuk_prl2009,gall_kolodka_prb2010,trojnar_korkusinski_prl2011,besombes_rossier_prb2012} 
The electrical control of the spin of manganese (Mn) ions in CdTe
quantum dots with a small electron population controlled by the gate
has been implemented.\cite{leger_besombes_prl2006}
The properties of CdTe quantum dots with magnetic impurities have been
extensively investigated
theoretically,\cite{hawrylak_grabowski_prb1991,efros_rashba_prl2001,govorov_prb2004,govorov_kalameitsev_prb2005,rossier_prb2006,qu_hawrylak_prl2005,qu_hawrylak_prl2006,nguyen_peeters_prb2008,reiter_kuhn_prl2009,trojnar_korkusinski_prl2011,oszwaldowski_zutic_prl2011}
including the theory of Coulomb blockade and capacitance
spectroscopy,\cite{efros_rashba_prl2001,qu_hawrylak_prl2005,rossier_aguado_prl2007}
cyclotron resonance,\cite{nguyen_peeters_prb2008} and photoluminescence (PL).\cite{govorov_prb2004,rossier_prb2006,cheng_hawrylak_epl2008,trojnar_korkusinski_prl2011,trojnar_korkusinski_prb2012}
The optical properties of carriers confined in III-V quantum
wells\cite{gazoto_brasil_apl2011} and quantum
dots\cite{kudelski_lemaitre_prl2007} containing Mn ions have also been
investigated.  

The electronic properties of quantum dots containing $N_e$ electrons
and a single localized spin have been investigated
theoretically.\cite{qu_hawrylak_prl2005,govorov_prb2004} 
It was shown that the electron spin can be controlled by controlling
the number $N_e$
(see Refs.~\onlinecite{qu_hawrylak_prl2005,cheng_hawrylak_epl2008}). 
For closed electronic shells the total spin is zero and electrons are
decoupled from the  Mn spin, while for half-filled shells the electron
spin is maximized and the coupling to the Mn spin is
strongest.\cite{qu_hawrylak_prl2005,cheng_hawrylak_epl2008} 
The spin-singlet $N_e$-electron droplet coupled to the Mn spin 
gives insight into the Kondo effect in the interacting electron
system, and this coupling might potentially allow for direct detection
of the electron spin.   
Simultaneously, the optical properties of charged  QDs without
magnetic ions have been studied both numerically and
experimentally.\cite{wojs_hawrylak_prb1997,narvaez_hawrylak_prb2000,ediger_karrai_pss2006,dalacu_reimer_lpr2010}

Here we present a microscopic theory of the optical properties of
charged self-assembled quantum dots doped with a single magnetic Mn
ion as a function of number of electrons $N_e$.  
The single-particle electron and heavy-hole electronic shells are
described by states of a two-dimensional harmonic oscillator. 
The electron-electron, electron-hole Coulomb interactions, as well as
the short-range electron spin-Mn spin and hole spin-Mn spin contact
exchange interactions are included. 
The electronic states of the photoexcited $N_e+1$ electron- $1$ hole-
$1$ Mn complex ($X^{N_e -}+$Mn) and of the final $N_e$ electron-Mn
complex ($N_e+$Mn) are expanded in a finite number of configurations. 
The  full interacting initial and final state Hamiltonians are
diagonalized numerically.  
The emission spectra  as a function of photon energy are obtained from
Fermi's golden rule as a function of $N_e$. 
We show that the emission spectra  depend on the number $N_e$,
the position of Mn ion, the spin of the initial and final electronic
states, and the size of the QD measured by the number of confined
electronic  shells.  
We demonstrate that the emission spectra allow to establish
the number of electrons $N_e$ populating electronic shells and,
most importantly, to read the electronic spin through multiplets of
energy levels manifested in the number of emission lines.  
If the Mn ion is placed in the center of the QD, the  $p$-shell
electrons do not interact with it directly.
However, we show that in this case there exists an effective
electron-Mn interaction mediated by electron-electron interactions. 
This mechanism allows to detect the spin polarization of a
half-filled $p$ shell.  

The paper is organized as follows. 
Section II describes the microscopic model, electronic structure,
total spin, and the emission spectrum of the system of many electrons
and a hole in a QD doped with a single Mn atom.  
Section III summarizes the results of the calculations of the emission
spectra from nonmagnetic QDs as a function of the number of 
initial-state electrons, and discusses in detail the emission from a
magnetic QD containing $p$-shell electrons, i.e., $X^{2-}$ and
$X^{3-}$ complexes. 
In this section, we also compare the emission spectra of $X$-,
$X^{-}$-, $X^{2-}$- and $X^{3-}$-Mn complexes and discuss the
differences and similarities between them. At last we discuss 
the effects of Mn position on the $X^{2-}$-Mn PL spectrum.
Summary of the work is presented in Sec. IV.

\section{Model}
We model the confining potential of the QD in the effective-mass
approximation as a quasi-two-dimensional isotropic harmonic oscillator
(HO).\cite{hawrylak_prl1993,wojs_hawrylak_prb1996} 
Since the strain in the QDs results in the significant splitting
between the light- ($\tau=\pm 1/2$) and heavy-hole ($\tau=\pm 3/2$)
subbands \cite{kadantsev_zielinski_jap2010}, we retain only
heavy-holes  in this calculations. 
We define the single-particle basis for electrons (hole) in terms of
the eigenstates of the isotropic parabolic quantum dot with the
characteristic frequency $\omega_{e (h)}$. 
The basis states are denoted by $|i \sigma\rangle$ for the electron
and $|j\tau\rangle$ for the hole, where the complex index denotes a
set of the HO quantum numbers $i=\{n,m\}$, while $\sigma$ ($\tau$) is
a spin $z$-projection of the particle. 
Each single-particle state has an angular momentum of $L_e=m-n$ for
the electron and $L_h=n-m$ for the hole.  
The energies of these single-particle states are given by $E_{i}^{e (h)}=\omega_{e (h)}(n + m + 1)$.

We measure energy in units of the effective Rydberg,
$\mbox{Ry}^{*} = m^{*} e^{4}/2\epsilon^{2} \hbar^{2}$,  
and length in the units of the effective Bohr radius, $a_{0}^{*} =
\epsilon \hbar^{2} /m^{*} e^{4}$, 
where $e$ is the electron charge, $\hbar$ is the reduced Planck
constant, $m^{*}$ is the effective mass of the electron, while
$\epsilon$ is the dielectric constant of the material.  

The Hamiltonian of the confined, interacting $N_e+1$ electrons and a
valence hole interacting with the spin of the magnetic impurity can be
written in the second quantization language
as\cite{wojs_hawrylak_prb1997,cheng_hawrylak_epl2008}  
\begin{align} \label{eq1}
H &=  \sum_{i,\sigma} E_{i,\sigma}^{e} c_{i,\sigma}^{\dagger} c_{i,\sigma} + \frac{1}{2}\sum_{\substack{i,j,k,l \\ \sigma, \sigma^{\prime} }} 
\langle i,j |V_{ee}| k,l \rangle c_{i,\sigma}^{\dagger} c_{j,\sigma^{\prime} }^{\dagger}c_{k,\sigma^{\prime} }c_{l,\sigma} \nonumber \\ 
+&\sum_{i,\tau} E_{i,\tau}^{h} h_{i,\tau}^{\dagger} h_{i,\tau} - \sum_{\substack{i,j,k,l \\ \sigma, \tau}} \langle i,j |V_{eh}| k,l \rangle 
c_{i,\sigma}^{\dagger} h_{j,\tau}^{\dagger}h_{k,\tau}c_{l,\sigma} \\ 
-&\sum_{i,j} \frac{J_{i,j}^{e}(R)}{2}\left[ \left(c_{i,\uparrow}^{\dagger} c_{j,\uparrow} - c_{i,\downarrow}^{\dagger} c_{j,\downarrow}
\right)M_{z} +  c_{i,\downarrow}^{\dagger} c_{j,\uparrow}M^{+}  \right.  \nonumber \\
& \left. + c_{i,\uparrow}^{\dagger} c_{j,\downarrow}M^{-} \right] +\sum_{i,j} \frac{3J_{i,j}^{h}(R)}{2}\left(h_{i,\Uparrow}^{\dagger} h_{j,\Uparrow} 
- h_{i,\Downarrow}^{\dagger} h_{j,\Downarrow} \nonumber\right)M_{z},
\end{align}
where $c_{i,\sigma}^{\dagger}$ ($h_{i,\tau}^{\dagger}$) creates an
electron (hole) on the orbital $i$ with spin $\sigma$ ($\tau$).  
The first two terms of the Hamiltonian are the electron kinetic energy
and the electron-electron Coulomb interaction (e-e). 
The next two terms describe the hole kinetic energy and the Coulomb
interaction between the hole and all electrons.  
The fifth term, describing the short-range electron-Mn (e-Mn)
interaction,\cite{qu_hawrylak_prl2005} consist of two types of terms.
The first one is the Ising interaction, which conserves the spin of
both the electron and the Mn.
The second and third terms of the e-Mn Hamiltonian allow for the
simultaneous flip of the electron and Mn spins in such a way as to
conserve $M_z+\sigma$.    
The last term is the anisotropic heavy-hole-Mn spin
interaction,\cite{leger_besombes_prb2005,govorov_prb2004} which
describes a scattering of the hole from $i$ to $j$ single-particle
states by a Mn ion at the position $R$. 
The e-Mn and hole-Mn (h-Mn) interaction is proportional to the \textit{s(p)-d}
exchange matrix elements, $J_{i,j}^{e (h)}(R) = J_{C}^{e (h)} 
\phi_{i}^{*}(R)\phi_{j}(R)$, 
where $\phi_{i} (R)$ is the value of the HO wavefunction at the
position $R$, while $J_{C}^{e (h)} = 2J_{s(p)-d}/d$. 
$J_{s(p)-d}$ is the bulk exchange contact interaction parameter, while
$d$ is the height of the QD.
As $J_{i,j}^{e (h)}(R)$ depends on the position $R$ of the Mn ion in
the QD, by changing $R$ one can control the strength of the e-Mn
interaction.\cite{qu_hawrylak_prl2005}

The many-particle wave function is expanded in the basis of the
configurations of $N_e+1$ electrons and a hole 
$|\nu_{i}\rangle =  
|i_{1\uparrow}, i_{2\uparrow},\ldots, i_{N\uparrow}\rangle|j_{1\downarrow},j_{2\downarrow},\ldots,j_{N\downarrow}\rangle|k\rangle|M_{z}\rangle$, 
where 
$|j_{1\sigma}, j_{2\sigma},\ldots, j_{N\sigma}\rangle = c_{j_{1},\sigma}^{\dagger}c_{j_{2},\sigma}^{\dagger}, \ldots, 
c_{j_{N_\sigma}}^{\dagger}|0\rangle$ is a state of $N_{\sigma}$
electrons, each with spin $\sigma$, while $|k\rangle =
h_{k,\tau}^{\dagger}|0\rangle$ is the hole state. 
$|M_{z}\rangle$ denotes all possible spin states of the Mn ion, $M_{z}
= -5/2, \ldots,5/2$, while $|0\rangle$ denotes the vacuum.  
The total number of electrons $N_e+1 = N_{\uparrow} + N_{\downarrow}$,
where $N_{\uparrow}$ and $N_{\downarrow}$ are the number of electrons
with spin up and spin down, respectively. 
After recombination and emission of a photon, we are left with $N_e$
electrons and the Mn ion. 
The final states of $N_e$ electrons $|\nu_{f}\rangle$ are built in a
similar way. 
The many-particle basis states are characterized by the total angular
momentum $L=\sum_{i=1}^{N_e+1} L_e^i +L_h$ as well as an electron and hole spins $s_z=\sum_{i=1}^{N_e+1} \sigma_i$ and $\tau$ or $L=\sum_{i=1}^{N_e}
L_e^i$ and $s_z=\sum_{i=1}^{N_e} \sigma_i$ for the initial and final states, respectively.

Having obtained the initial and final states, one can calculate the
circularly polarized emission spectra from the Fermi's golden rule:
\begin{equation} \label{eqPL}
E(\omega) = \sum_{f} P_i |\langle \nu_{f}|\mathcal{P}|\nu_{i} \rangle|^{2} \delta (\mathcal{E}_{i} - \mathcal{E}_{f} - \omega),
\end{equation}
where $\omega$ is the photon energy, while $\mathcal{E}_{i}$ and
$\mathcal{E}_{f}$ are the energies of the initial and final states,
respectively. 
The coefficient $P_i$ is the probability of thermal occupation of the
initial state $|\nu_{i} \rangle$,
$P_i=\exp\left(-\mathcal{E}_{i}/kT\right)/P_{SUM}$, with
$P_{SUM}=\sum_i\exp\left(-\mathcal{E}_{i}/kT\right)$.  
The interband polarization operator $\mathcal{P} = \sum_{kl} \langle
k|l \rangle c_{k}h_{l}$ removes one electron-hole pair from the
initial state. 
The optical selection rules \cite{wojs_hawrylak_prb1997} are defined
by the overlap $\langle k|l \rangle$ between the electron and hole
orbitals.  

\section{Electronic structure and emission spectra of charged magnetic dots}
 
In this section, we present the results of numerical calculation of the
emission spectra of multiply charged QDs doped with a single Mn ion.  
Recent experiments and theory
\cite{trojnar_korkusinski_prl2011,trojnar_korkusinski_prb2013}
indicate that in the CdTe quantum dots there are at least three
confined single-particle shells, $s$, $p$, and $d$, and the presence of
the $d$ shell can give rise to new effects, such as the quantum
interference (QI).\cite{trojnar_korkusinski_prl2011} 
When Mn is in the center of the QD ($R=0$), only those electrons that
occupy the zero angular momentum orbitals are coupled with it.
In the presence of three shells in the QD, there are two zero angular
momentum orbitals, one in the $s$ and one in the $d$ shell.  

The calculations are carried out with the following parameters:
$Ry^{*}=12.11$ meV,  $a_{0}^{*}=5.61$ nm, for CdTe with the
dielectric constant $\epsilon=10.6$. 
The electron and hole effective masses are $m^{*}=0.1m_{0}$ and
$m_{h}=4m^{*}$, respectively, with $m_{0}$ being the free-electron
mass. 
The electron characteristic frequency $\omega_{e} = 1.98$ $\mbox{Ry}^{*}$ and
$\omega_{h} = \omega_{e}/4$.   
The constant scaling the exchange contact interaction in the bulk
CdTe for electrons is $J_{s-d} = 15$ meV$\cdot$nm$^{3}$, and for holes
is $J_{p-d} = 60$ meV$\cdot$ nm$^{3}$, while the height of the QD
$d=2$ nm.

As already mentioned, first we present the emission spectra for a 
nonmagnetic QD for $N_e=0$ to $6$.
In the case of $N_e=6$, both $s$ and $p$ shells of the QD are filled. 
After that we investigate in detail the PL
of $X^{2-}$ and $X^{3-}$ complex for the QD doped with a single Mn
atom in its center.  
Lastly, we compare the emission spectra from $X$, $X^{-}$,
$X^{2-}$ and $X^{3-}$ complexes confined in a magnetic QD, and discuss
their features. 

\subsection{Emission from a nonmagnetic charged quantum dot}
\begin{figure}
\begin{center}
\includegraphics[width=0.48\textwidth]{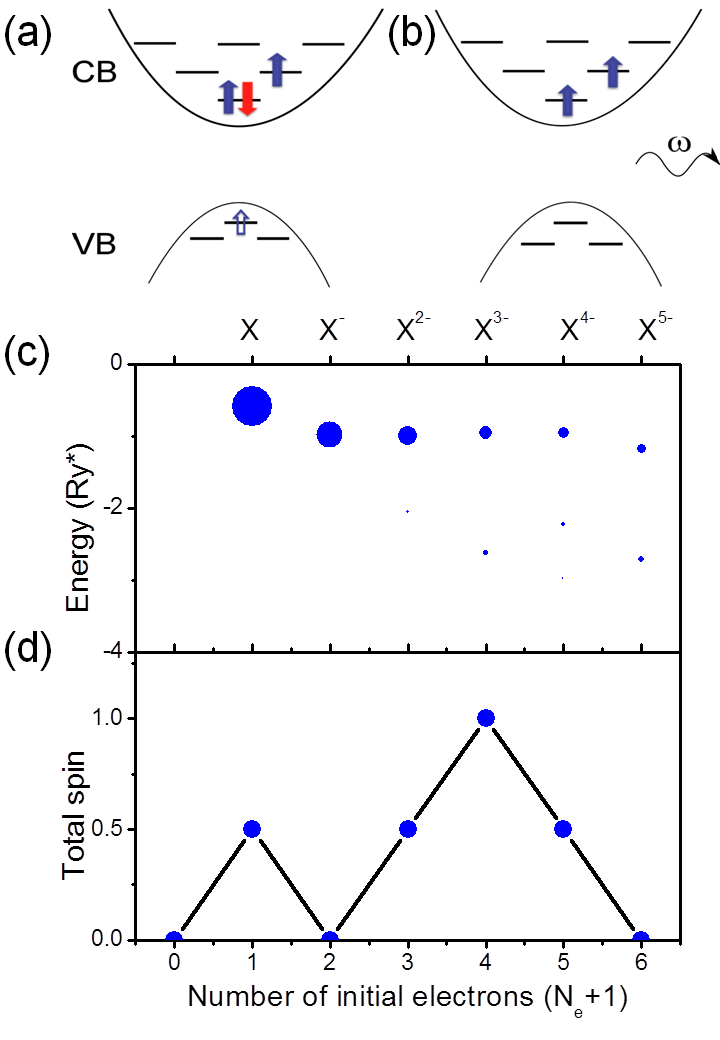}
\caption{(Color online) A schematic representation of (a) the 
ground state of an $X^{2-}$ complex, formed by three electrons and a
spin-up hole, and (b) the final state after recombination of electron-hole
pair from the configuration (a).
(c) Emission spectrum of a nonmagnetic QD calculated in the
$\sigma_{+}$ polarization as a function of the number of electrons in
the initial state. 
It is assumed that the  probability $P_i$ of occupation each of the
degenerate initial states is equal.   
(d) Total spin of electrons in the ground state of the initial system
as a function of the number of confined electrons.
\label{fig1}}
\end{center}
\end{figure} 

Figure~\ref{fig1}(a) schematically shows a ground state of the $X^{2-}$
complex, composed of three electrons in the conduction band (CB) and a
hole in the valence band (VB).  
Two electrons and the hole occupy the $s$ shell, while the third
electron is in the $p$ shell.  
After the electron-hole recombination from the $X^{2-}$ complex, the
final state is formed by two electrons in the CB and a photon with
energy $\omega$, as shown in Fig.~\ref{fig1}(b). 
The remaining two electrons can be either in a triplet or in a singlet
spin configuration, which have the same kinetic energy but are split
by the e-e exchange Coulomb interaction.\cite{wojs_hawrylak_prb1997}

Figure~\ref{fig1}(c) shows the evolution of the recombination spectrum
in $\sigma_{+}$ polarization as a function of the number of electrons
in the initial (photoexcited) state.
The area of the circles is proportional to the intensity of individual
transitions. 
The emission is symmetric with respect to the hole spin, so the
$\sigma_{-}$ polarization spectrum is exactly the same. 

For doubly ($X^{2-}$) and higher-charged exciton states the emission
peak splits into two or more. 
The splitting in the emission spectra originates from the splitting of
the final state as discussed above and in
Ref. \onlinecite{wojs_hawrylak_prb1997}.   
From the  emission spectra of nonmagnetic QDs, one can
not draw conclusion about the spin of $N_e$ electrons left after the
electron-hole pair recombination. 

Figure~\ref{fig1}(d) illustrates the total electron spin of the initial
ground state as a function of number of electron in this state. 
The QD is filled obeying the QD Hund's rule.\cite{wojs_hawrylak_prb1997} 
Until each shell is half-filled, subsequent electrons are added with
the same spin, increasing the total spin of this shell. 
After the half-filling is reached, electrons are added
with opposite spin, which results in the spin zero of a completely
filled shell. 
As the $p$ shell is being filled, the  maximum spin is
reached when there are four electrons in the QD, while in the
presence of six electrons, both $s$ and $p$ shells are filled, and the
spin of electrons is equal to zero. 

\subsection {$X$-Mn and $X^-$-Mn complexes}

The emission spectra of both the exciton ($X$) and negatively charged
exciton $X^-$ interacting with the spin of the Mn ion have been
described
previously.\cite{besombes_leger_prl2004,leger_besombes_prl2006,trojnar_korkusinski_prl2011}
Here we briefly summarize these results.

The ground state (GS) of the exciton-Mn system can be approximated by the
configuration in which the electron and the hole occupy their respective
$s$ shell. 
The final states left after the electron-hole pair recombination are
the degenerate states $|M_z\rangle$ of Mn.
Thus, the emission spectrum from $X$-Mn complex consists of six
emission lines (one for each $M_z$). 
The splitting between these lines corresponds directly to the
splittings between $X$-Mn states, approximated by 
$1/2\left(3J_{ss}^h+J_{ss}^e\right)$ in the symmetric
QDs.\cite{besombes_leger_prl2004} 
Since the h-Mn exchange constant $J_{ss}^h$ is four times greater
than the e-Mn exchange constant $J_{ss}^e$, the splittings are
dominated by the h-Mn
interaction.\cite{besombes_leger_prl2004,trojnar_korkusinski_prl2011}

The GS of a negatively charged exciton $X^{-}$ interacting with Mn
consists of two electrons and one hole, all occupying the
single-particle $s$ shell. 
The two electrons are in the singlet state, which prevents them from
interacting with Mn. 
$X^{-}$ interacts with Mn only through the h-Mn Ising Hamiltonian,
which splits the otherwise degenerate $X^{-}$-Mn into six levels
similarly to the $X$-Mn case.  
However, in contrast to the $X$-Mn complex, there are two final states
of the one e-Mn system with one electron on the $s$ shell
interacting with the Mn ion. 
These two e-Mn states have $J=S+M=2$ or $J=3$ and are
split by the $e$-Mn interaction. Since the emission from the initial state with $M_{z}=5/2$ to the
final state with $J=2$ is forbidden, the emission spectrum of the
$X^{-}$-Mn has eleven lines arranged into six groups.\cite{leger_besombes_prl2006} 
The emission spectra of the $X$-Mn and $X^-$-Mn complexes will be
shown in greater detail later on.

\subsection{$X^{-2}$-Mn complex}

Here we present the emission spectra from $X^{2-}$-Mn complex confined
in our magnetic QD with a single Mn ion in its center. 
We begin with a detailed description of the electronic structure of
both initial and final states and then discuss the calculated emission
spectrum.  
\begin{figure}
\begin{center}
\includegraphics[width=0.48\textwidth]{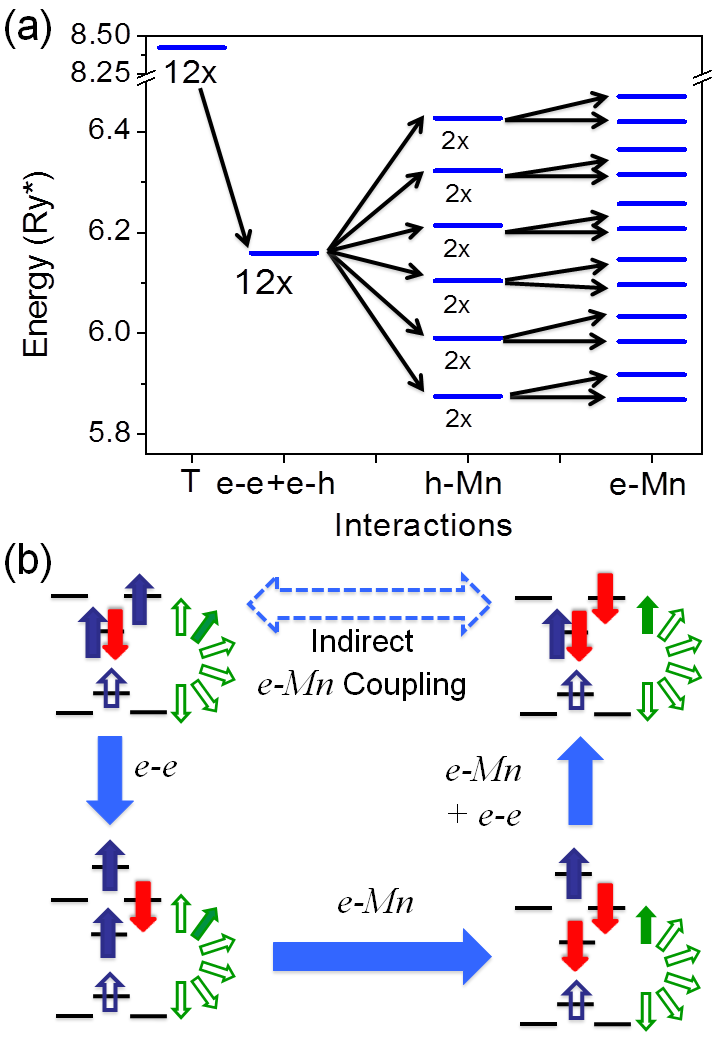}
\caption{(Color online) 
(a)  The ground-state energy of the $X^{2-}$-Mn system as a function
of interactions (subsequent terms added towards the right-hand
side of the panel).
The degeneracy of the energy levels is marked. 
The splitting in the ground state caused by e-Mn interaction is not
to scale. 
(b) The coupling scheme between the $|X^{2-}, s_z=1/2\rangle$ and
$|X^{2-}, s_z=-1/2\rangle$ configurations. 
Filled arrows indicate the direct coupling, while 
the dashed arrow indicates an indirect coupling.  \label{fig2}}
\end{center}
\end{figure}
\subsubsection{Initial state}
The ground state of the $X^{-2}$ complex confined in a nonmagnetic QD
can be approximated by the lowest-energy configuration shown
schematically in Fig.~\ref{fig1}(a).  
The $X^{-2}$ GS is fourfold degenerate: twice with respect to the spin
of the electron in the $p$ shell, and twice with respect to its
angular momentum ($L=\pm 1$).  
Since the total angular momentum $L$ is a good quantum number, the
analysis will be carried out in the $L=1$ subspace. 
The double degeneracy of the GS of $X^{2-}$ complex due to the spin
persists even in the presence of e-e and e-h interactions. 
The two main configurations of the $X^{2-}$ complex are: 
$|X^{2-}, s_z=1/2\rangle =
c^+_{s\uparrow}c^+_{p\uparrow}c^+_{s\downarrow}h^+_{s\Uparrow}|0\rangle$
and  
$|X^{2-}, s_z=-1/2\rangle = c^+_{s\uparrow}c^+_{s\downarrow}c^+_{p\downarrow}h^+_{s\Uparrow}|0\rangle$.
These two states do not interact with each other since they have
different spin projections $s_z$.
In a QD with three confined single-particle shells, one can construct
198 different configurations of $X^{2-}$ complex with $L=1$ which
interact either with $|X^{2-}, s_z=1/2\rangle$ or with $ |X^{2-},
s_z=-1/2\rangle$ via e-e and e-h Coulomb interactions. 
In a magnetic QD with Mn in its center, the angular momentum is
conserved, and the total number of the $X^{2-}$-Mn configurations
increases $(2M+1)=6$ times, reaching 1188 configurations in the $L=1$
subspace. 

Figure~\ref{fig2}(a) shows the evolution of the $X^{2-}$-Mn low-lying
energy spectrum with inclusion of interactions.
The first column shows calculations with only kinetic energy $T$,
second after we include e-e and e-h Coulomb interactions, third with
h-Mn interaction added, and finally fourth includes e-Mn
interaction.   
In the absence of any interactions the GS is twelvefold degenerate,
twice due to electron spin and six times due to Mn spin
orientations. 
This degeneracy does not change after inclusion of the e-e and e-h
Coulomb interactions, however the energy of the complex decreases.  
After addition of the h-Mn Ising-type interaction, Eq. (\ref{eq1}),
the ground state of the $X^{2-}$-Mn complex splits into six doubly 
degenerate levels.  
Since none of the $|X^{2-}, s_z=1/2\rangle$ and $|X^{2-},
s_z=-1/2\rangle$ configurations interact directly with the Mn via 
e-Mn interaction, $\langle X^{2-}, s_z=1/2|H_{e-Mn}|X^{2-},
s_z=1/2\rangle=\langle X^{2-}, s_z=-1/2|H_{e-Mn}|X^{2-},
s_z=-1/2\rangle=0$, one expects no change in the energy spectra of the
$X^{2-}$-Mn complex after inclusion of the e-Mn interaction as in Eq. (\ref{eq1}). 
However, addition of the e-Mn interaction leads to further splitting
of the $X^{2-}$-Mn energy levels as shown in Fig.~\ref{fig2}(a). 
Since the e-Mn coupling is smaller than the h-Mn coupling, the
resulting splitting is magnified out of scale in order to visualize
the effect.   
The origin of this splitting is in the indirect coupling between the
$p$-shell electrons and Mn, mediated by e-e Coulomb interaction.  

Figure~\ref{fig2}(b) shows the coupling scheme among seemingly
noninteracting configurations of $X^{2-}$ complex and Mn ion.   
In order to simplify the discussion we focus only on the splitting
caused by the e-Mn interaction. 
The filled arrows indicate a direct coupling between the $X^{2-}$-Mn
configurations, while the dashed arrow represents an indirect coupling.  

Let us start form the $|X^{2-}, s_z=1/2\rangle \otimes |M_z\rangle$
configuration, see Fig.~\ref{fig2}(b) top left. 
This configuration is coupled via e-e Coulomb interaction with an
excited configuration (with the same $s_{z}$ and $M_{z}$) with a
spin-down electron in the $p$ shell, and two unpaired electrons in the
$s$ and $d$ shells.  
The e-Mn interaction can flip the spin of the spin-up electron in the
$s$ shell with simultaneous flip of the Mn spin to $M_{z}+1$, as
illustrated in the bottom of Fig.~\ref{fig2}(b).  
This excited state with $s_{z}=-1/2$ and $M_{z}+1$ is coupled with the
$|X^{2-}, s_z=-1/2\rangle \otimes |M_z+1\rangle$ via the e-e Coulomb
interaction as well as the e-Mn interaction ($J_{sd}(0)\ne 0$).  
 
One can replace the coupling scheme between $|X^{2-}, s_z=1/2\rangle
\otimes |M_z\rangle$ and $|X^{2-}, s_z=-1/2\rangle \otimes
|M_z+1\rangle$ configurations presented in Fig.~\ref{fig2}(b) by
filled arrows  with a single dashed arrow, representing the indirect
coupling. 
Effectively, one can look at this coupling as that of the $p$-shell
electrons to a Mn ion, mediated by e-e and e-Mn interactions.   

\subsubsection{Final state}

The final state left after the electron-hole recombination from the
$X^{2-}$-Mn complex is composed of one electron in the $s$ shell and
one in the $p$ shell, as illustrated in Fig.~\ref{fig1}(b). 
This state has the same total angular momentum as the initial state,
namely, $L=1$. The remaining electrons can be in a triplet state with total spin
$S=1$, or in a singlet state with $S=0$. 
These states have the same kinetic energy, therefore in the absence of
any interaction they are 24 times degenerate, four times due to
electron spin and six times due to Mn spin.  
\begin{figure}
\begin{center}
\includegraphics[width=0.48\textwidth]{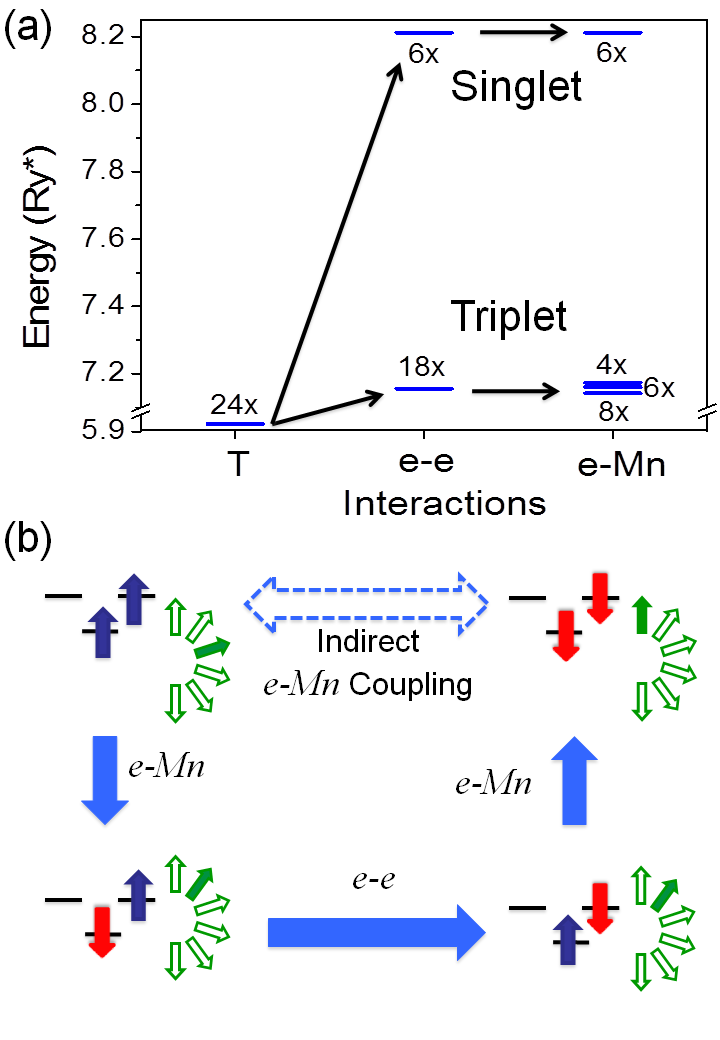}
\caption{(Color online) (a) Evolution of the final-state energy as a function of interactions, with marked degeneracy of the energy
  levels. 
(b) The coupling scheme between the final state configurations. Types
of arrows have the same meaning as in Fig.~\ref{fig2}(b).   
\label{fig3}}
\end{center}
\end{figure}
 
Figure~\ref{fig3}(a) shows the evolution of the energy of $X^{2-}$ final
states with the inclusion of different types of interactions.  
The e-e Coulomb interaction splits the triplet and the singlet
states. 
The triplet is 18 times degenerate while the singlet is sixfold
degenerate. 
Addition of the e-Mn interaction leads to the splitting of the triplet
state into three levels with eight-, six- and fourfold degeneracy. 
The triplet state of the pair of electrons on the $s$ and $p$ shell
experiences the same kind of splitting as a particle with $S=1$
interacting with Mn via ferromagnetic interactions.
At the same time, the singlet remains six times degenerate.

To understand the fine structure of the triplet state, let us analyze
how the electrons interact with Mn. Figure~\ref{fig3}(b) shows the coupling 
scheme between two pairs of $s$-$p$ electrons: $|s_z=1\rangle \otimes|M_z\rangle$ 
and $|s_z=-1\rangle\otimes|M_z+2\rangle$ in the presence of the Mn ion. 
Again, the filled arrows indicate a direct coupling between the
configurations, while the dashed arrow indicates an indirect coupling. 
Let us start from the configuration $|s_z=1\rangle
\otimes|M_z\rangle$, Fig.~\ref{fig3}(b), top left. 
The e-Mn spin-flip interaction couples it with the configuration with
$s_{z}=0$ and $M_{z}+1$, where the electron with the spin $s_z=-1/2$
is on the $s$ shell. 
This configuration is coupled by the e-e Coulomb interaction 
with a configuration with $s_{z}=0$ and $M_{z}+1$, but the spin down
electron occupying the $p$ shell. 
Again, the e-Mn interaction couples the former configuration with the
$|s_z=-1\rangle\otimes|M_z+2\rangle$, through the e-Mn spin-flip
process.  
Effectively, the coupling scheme shown by the filled arrows can be
replaced by the dashed arrow, representing the indirect coupling
between configurations $|s_z=1\rangle\otimes|M_z\rangle$ and
$|s_z=-1\rangle\otimes|M_z+2\rangle$. 
From this coupling scheme, one can conclude that indeed a pair of $s$-$p$
electrons interacts with Mn ion in the same way as a spin $S=1$
particle, with three possible spin projections $S_z=-1,0,1$, changing
the spin of Mn from $|M_z\rangle$ to $|M_z+2\rangle$, with
simultaneous change of its spin from $S_z=1$ to $S_z=-1$. 
The difference is that in a spin Hamiltonian both spins ($S=1$ and
$M_z=5/2$) interact directly, while in our model, the coupling with the
$p$-shell electron is indirect.    

\subsubsection{Emission spectrum}

Having obtained the $X^{2-}$ initial and final eigenvalues and
eigenstates, we calculate the emission spectrum of $X^{2-}$-Mn
complex. 
The spectrum in the $\sigma^{+}$ polarization is calculated at
temperature in which a thermal population of the twelve lowest
$X^{2-}$-Mn states is equal [$P_i=1$ in Eq. \ref{eqPL}]. 
The $\sigma^{+}$ polarized light is emitted due to the recombination
of the spin-up hole and a spin-down electron from the
initial $X^{2-}$-Mn state. 
Both total angular momentum of the electronic state as well as $M_z$
are conserved in the recombination process.  

\begin{figure}
\begin{center}
\includegraphics[width=0.48\textwidth]{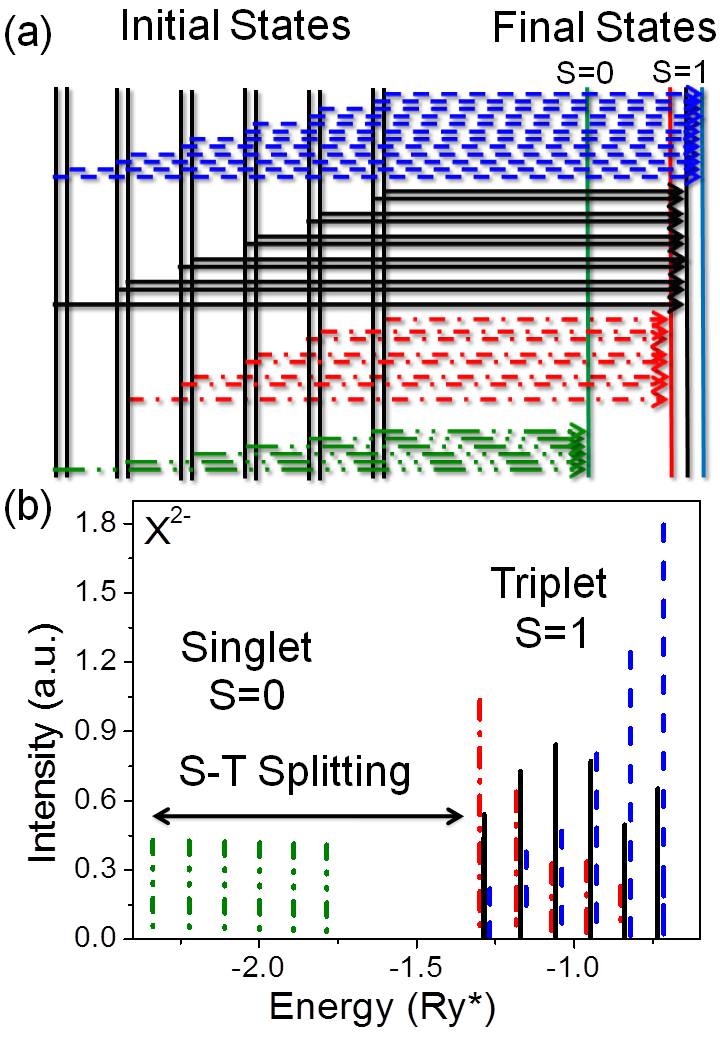}
\caption{(Color online) (a) The energy levels of initial and final
  states  with allowed transitions between them indicated by the
  arrows. 
The dashed, solid,  and dashed-dot arrows denote the transitions to
the $S=1$ electron states, while the dashed-double-dotted arrow
denotes the transitions to the $S=0$ final electron states. 
(b) Emission spectrum of $X^{2-}$-Mn complex calculated in the
$\sigma_+$ polarization as a function of the photon
energy. \label{fig4}} 
\end{center}
\end{figure}

Figure~\ref{fig4}(a) schematically shows the energy levels of both
initial and final states.
The dashed, solid, and dashed-dotted arrows indicate the optically
active transition from all twelve initial states to the triplet final
states, split into three levels with degeneracy eight, six, and four.  
The dashed-double-dotted arrows present six optically active
transitions from the initial states with electron spin $s_z=-1/2$ to the
sixfold degenerate singlet state.  

The emission spectrum from the $X^{2-}$-Mn complex is shown in
Fig.~\ref{fig4}(b), with the colors and styles of lines corresponding
to the styles of arrows in Fig.~\ref{fig4}(a). 
The emission spectrum consists of two groups of transitions [as in a
QD without Mn in Fig.~\ref{fig1}(c)]: the lower lying transitions
(six dashed-double-dotted lines) correspond to the transition to the
final state with the two electrons in a singlet state ($S=0$), while
the higher-lying group corresponds to the transitions to their triplet
state ($S=1$).  

Let us start the discussion by analyzing the lower-energy part of the
spectrum presented by dashed-double-dotted peaks. 
Since the final state$-$in this case an electron singlet$-$is
degenerate, the splitting between the emission lines corresponds to
the splitting between the initial states due to the hole-Mn
interaction.   
One could expect that each of the lines is in fact a doublet, with the
small splitting between them due to the e-e and e-Mn interactions,
however only one of them is bright: only states with $s_z=-1/2$ can
have an electron singlet as a final state.  
The existence of the {\it six} lines in the emission confirms the
previous hypothesis of the $s$-$p$ electron pair behavior as a $S=0$
particle. 

In the higher-energy part of the emission spectra
(see Fig.~\ref{fig4}(b)), the transitions to the final states of electrons
in a triplet states are presented. 
This part of emission spectra consists of six groups of peaks split
by the h-Mn interaction in the initial state. 
Five of the main peaks are then further split into three due to the e-e
and e-Mn-induced splitting in the final state and correspond to
different final states. 
The highest-energy  main peak is split into two, since the transition
from the highest energy state of the $X^{2-}$-Mn complex to the
fourfold-degenerate final state is dark.   
Therefore, by looking at the energy difference between two consecutive
peaks, within the same main peak, i.e., the solid and dashed-dot
peaks, one can obtain the effective splitting of the final state,
which depends on e-e and e-Mn interactions.  
Each of these peaks can be further split, reflecting the e-Mn induced
splitting in the initial state giving total of 31 optically active
transitions, however this splitting is not visible on the scale of
Fig.~\ref{fig4}(b). 

\subsection{$X^{3-}$-Mn complex}

The simplest system allowing to study the behavior of electrons in a
half-filled shell is the $X^{3-}$ complex. 
In the  GS of this complex, the two electrons in the $p$ shell are in
a triplet state $S=1$, which makes them a good candidate to interact
with the Mn spin.
Indeed, as we have shown previously, there exists an effective
interaction between the $p$-shell and the Mn spin mediated by
the e-e Coulomb interactions.
Here we describe the electronic properties of the initial and final
states of the $X^{3-}$ complex  and its emission spectrum. 
\subsubsection{Initial state}
\begin{figure}
\begin{center}
\includegraphics[width=0.48\textwidth]{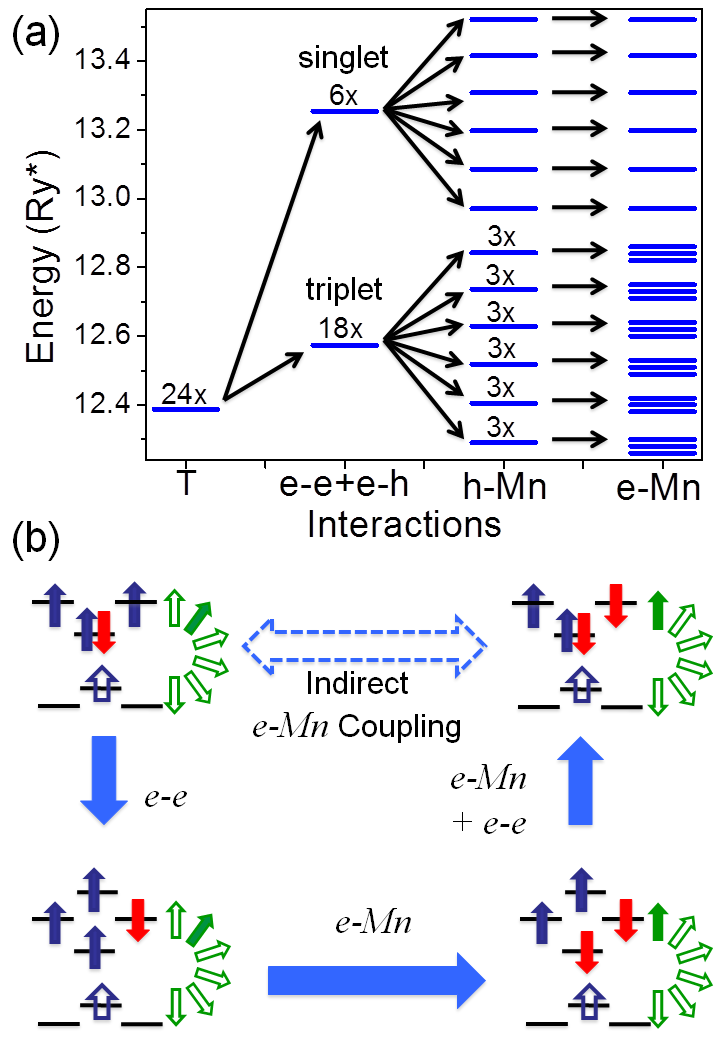}
\caption{(Color online) 
(a) The ground-state energy of  $X^{3-}$-Mn complex as function of
interactions, with marked degeneracy of the energy levels. 
The splitting in the GS caused by e-Mn interaction is out of scale. 
(b) The coupling scheme between $X^{3-}$-Mn configurations with
different $s_{z}$. Types of arrows have the same meaning as in
Fig.~\ref{fig2}(b).  
\label{fig5}}
\end{center}
\end{figure}

The GS of the $X^{3-}$ is composed of four electrons and one hole.
As previously, we focus only on one subspace, with the hole spin
projection $\tau=3/2$. 
The lowest-energy configurations of the $X^{3-}$ complex have total
angular momentum equal to zero, and are fourfold degenerate in the
absence of e-e Coulomb interactions. 
The degeneracy is due to the four possible spin configurations of the
two unpaired electrons in the $p$ shell: they can be either spin
polarized ($s_z=\pm 1$) or with $s_z=0$, creating either a singlet or a
triplet state. 

The total number of the $X^{3-}$-Mn configurations with the angular
momentum $L=0$ and  $\tau=3/2$ in a QD confining three single-particle shells is 2664.
The excited configurations of the $X^{3-}$-Mn complex play an
important role in mediating the interactions between the $p$-shell
electrons and Mn. 
Therefore, the $X^{3-}$ complex allows us to study the behavior of 
$S=1$ and $S=0$ spins interacting indirectly with Mn. 

Figure~\ref{fig5}(a) shows the evolution of the GS energy as we add
the interactions. 
With inclusion of the Coulomb interactions between the carriers, the
24-fold degenerate $X^{3-}$-Mn GS splits into two states: a
lower-lying triplet-Mn (18-fold degenerate) and a singlet-Mn (sixfold
degenerate).  
The h-Mn interaction breaks the Mn symmetry and splits both of these
manifolds into six levels. 
Each of the six triplet-Mn levels is still threefold degenerate.  
This degeneracy is lifted by e-Mn interaction.
This takes place only in the presence of the e-e Coulomb interaction
in the QDs containing at least three shells and it is another proof
that both of the $p$ electrons are coupled indirectly with the Mn ion.
The splitting of singlet-Mn state remains unchanged, but its energy is
lowered in relation to the system without e-Mn interaction.  
As the e-Mn splitting is smaller than the h-Mn splitting, it is out of
scale in Fig.~\ref{fig5}(a). 

Figure~\ref{fig5}(b) shows the coupling scheme between $X^{3-}$-Mn
configurations with different $s_{z}$.  
The configuration of $X^{3-}$-Mn with two spin-up electrons in the
$p$ shell and the Mn spin $M_{z}$ is coupled by e-e Coulomb interactions
with an excited configuration (with the same $M_{z}$) in which the
spin-down electron is scattered to the $p$ shell and the spin-up
electron is scattered to the $d$ shell [as shown in the bottom-left
panel of Fig.~\ref{fig5}(b)]. 
The e-Mn interaction can flip the spin of the electron occupying the
$s$ shell, with a simultaneous increase of the Mn spin by one (to the
state with $M_z+1$). 
Now, this configuration is coupled via e-e Coulomb interactions with a
low-energy $X^{3-}$-Mn configuration in which there is a spin-up and
spin-down electron in the $p$ shell and the Mn spin is in the state
$M_{z}+1$.  
Therefore, all initial states forming the GS manifold are indirectly
coupled via e-e and e-Mn interactions, as shown by the dashed arrow. 
This coupling breaks the triplet degeneracy, and can be again treated
as one induced by an indirect coupling between two $p$-shell electrons
and the Mn ion.   

\subsubsection{Final state}

The final state, left over after recombination of the spin down
electron with the spin up hole from the $X^{3-}$ complex,  is an {\it
 excited state} of the three electrons system (with $L=0$).
It is formed by one electron in the $s$ shell and two in the $p$
shell. 
However, since the configurations with two electrons in the $s$ shell
and one in the $d$ shell have the same kinetic energy as the
configurations mentioned before, they are strongly coupled by the e-e
Coulomb interaction.   

\begin{figure}
\begin{center}
\includegraphics[width=0.48\textwidth]{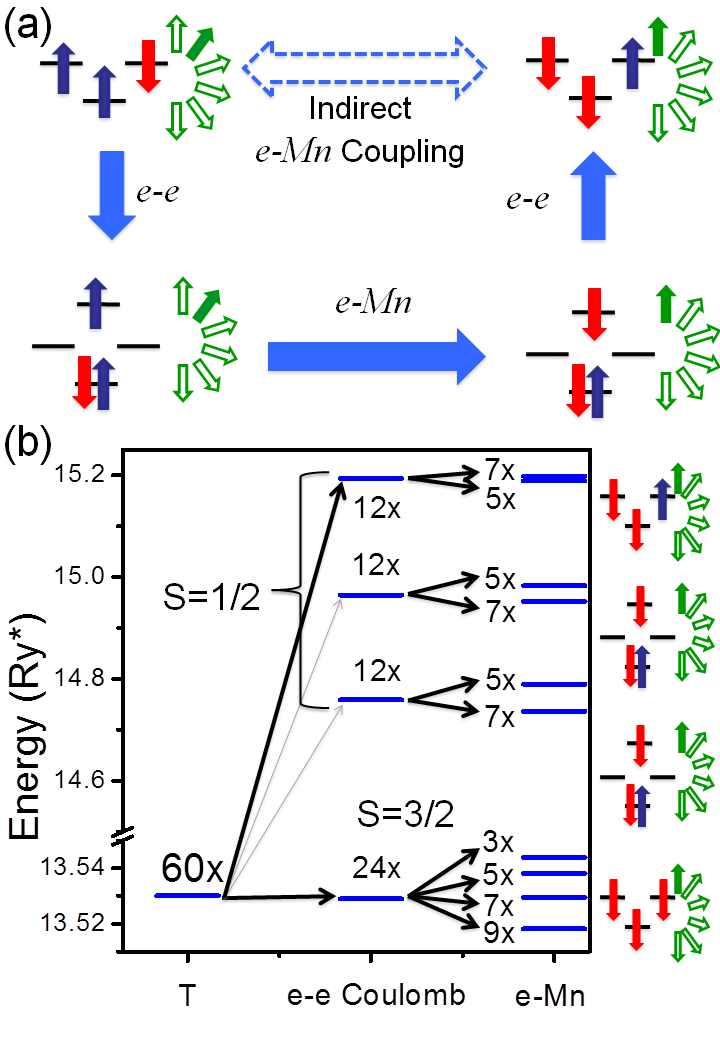}
\caption{(Color online) (a) The coupling scheme between the
  three-electron-Mn configurations with different $s_{z}$. Types of
  arrows have the same meaning as in Fig.~\ref{fig2}(b). 
(b) Evolution of the final-state energy with various interaction terms,
the numbers indicate the degeneracy of the energy levels. \label{fig6}} 
\end{center}
\end{figure}

Figure~\ref{fig6}(a) shows the coupling scheme between four
three-electron-Mn (3e-Mn) configurations with the same kinetic energy
but with different $s_{z}$. 
The coupling mechanism between the configurations is the same as that
explained in previous sections.  

Figure~\ref{fig6}(b) illustrates how the energies of the 3e-Mn system
evolve as interaction terms are added.  
In the absence of any interactions, all of the considered
configurations have the same energy.
This energy level is 60-fold degenerate, ten due to electron
configurations times six Mn spin orientation. 
Addition of the e-e Coulomb interaction splits the energy of the 3e-Mn
complex into four levels, with degeneracy 24, 12, 12, 12,
respectively.   
The lowest-energy electron state has total spin $S=3/2$, while all
higher energy levels correspond to $S=1/2$.  
It is important to notice that the two intermediate $S=1/2$ states are
not final states of the $X^{3-}$ complex recombination in a QD with or
without Mn.
It is so because they are built by the configurations with two
electrons in the $s$ shell and the third electron in the $d$ shell
mixed with the configurations with one electron on the $s$ shell and a
pair of electrons occupying the $p$ shell in a singlet state. 
At the same time the lowest and higher energy states are formed mostly
by configurations with only one electron in the $s$ shell and two in
the $p$ shell, allowing them to be final states in the recombination
of the $X^{3-}$(-Mn) complex. 

In the presence of the e-Mn interaction, the $S=3/2$ level splits into
four states and each of the $S=1/2$ levels into two.
We observe that the $S=3/2$ and two first $S=1/2$ three-electron
states are ferromagnetically coupled with Mn, since the degeneracy of
levels (related to total electron-Mn spin) decreases as a function of
energy. 
The ordering of states is different in the case of highest $S=1/2$
state which is antiferromagnetically coupled with Mn.  

\subsubsection{Emission spectrum}
\begin{figure}
\begin{center}
\includegraphics[width=0.48\textwidth]{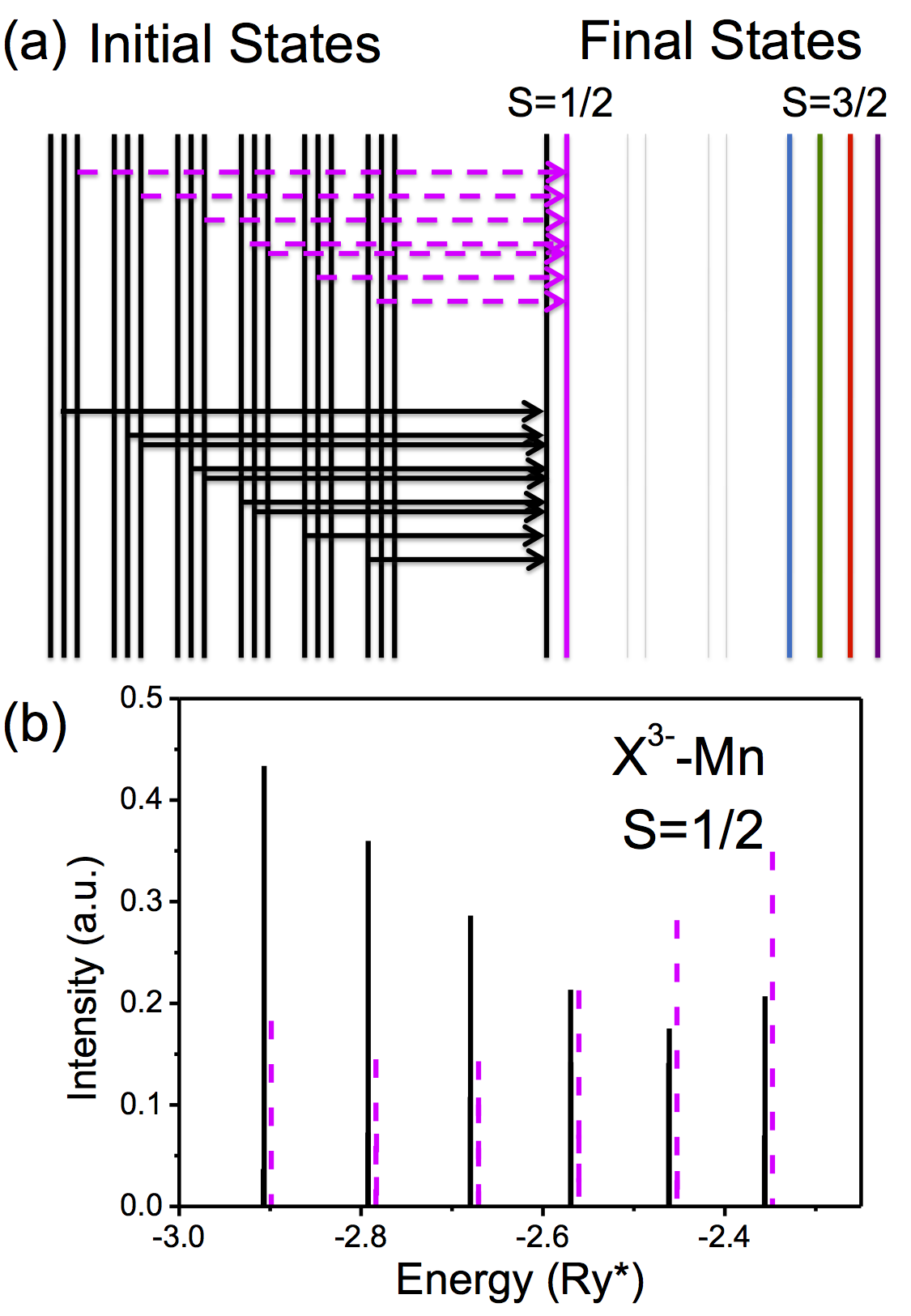}
\caption{(Color online) (a) Schematic representation of the
  $X^{3-}$-Mn initial and final states, with the allowed transitions
  to the final states with $S=1/2$ indicated by the dashed arrows.  
(b) The low-energy $X^{3-}$-Mn emission spectrum in $\sigma_+$
polarization. 
The styles of emission lines correspond to the styles of arrows (a).
\label{fig7a}}
\end{center}
\end{figure}

Here we investigate the emission from the equally populated initial
states of the $X^{3-}$-Mn complex with $S=1$ [18 lowest energy levels
in Fig.~\ref{fig5}(a)].  
The emission spectra from the $X^{3-}$-Mn complex consist of two main
groups of peaks corresponding to optical transitions from the initial
states to the $S=3/2$ and the highest of $S=1/2$ final states, in the
way resembling the emission from $X^{3-}$ in nonmagnetic QD
[ see Fig.~\ref{fig1}(b)]. 
These two types of transitions will be analyzed separately. 

Figure~\ref{fig7a}(a) shows the $X^{3-}$-Mn initial and final energy
levels, with the dashed arrows denoting the transitions to the highest
two final states with $S=1/2$.  
The emission spectrum from $X^{3-}$-Mn complex to these final states
consists of six groups of peaks and it is shown in
Fig.~\ref{fig7a}(b).  
The splitting into main six groups is caused by the h-Mn interaction in
the initial state, while the splitting into two peaks in each group is
due to the e-Mn interaction in the final state. 
The effects of the splitting  of the initial state induced by the e-Mn
interaction are not visible in this figure, but they cause further
splittings of emission lines.  
\begin{figure}
\begin{center}
\includegraphics[width=0.48\textwidth]{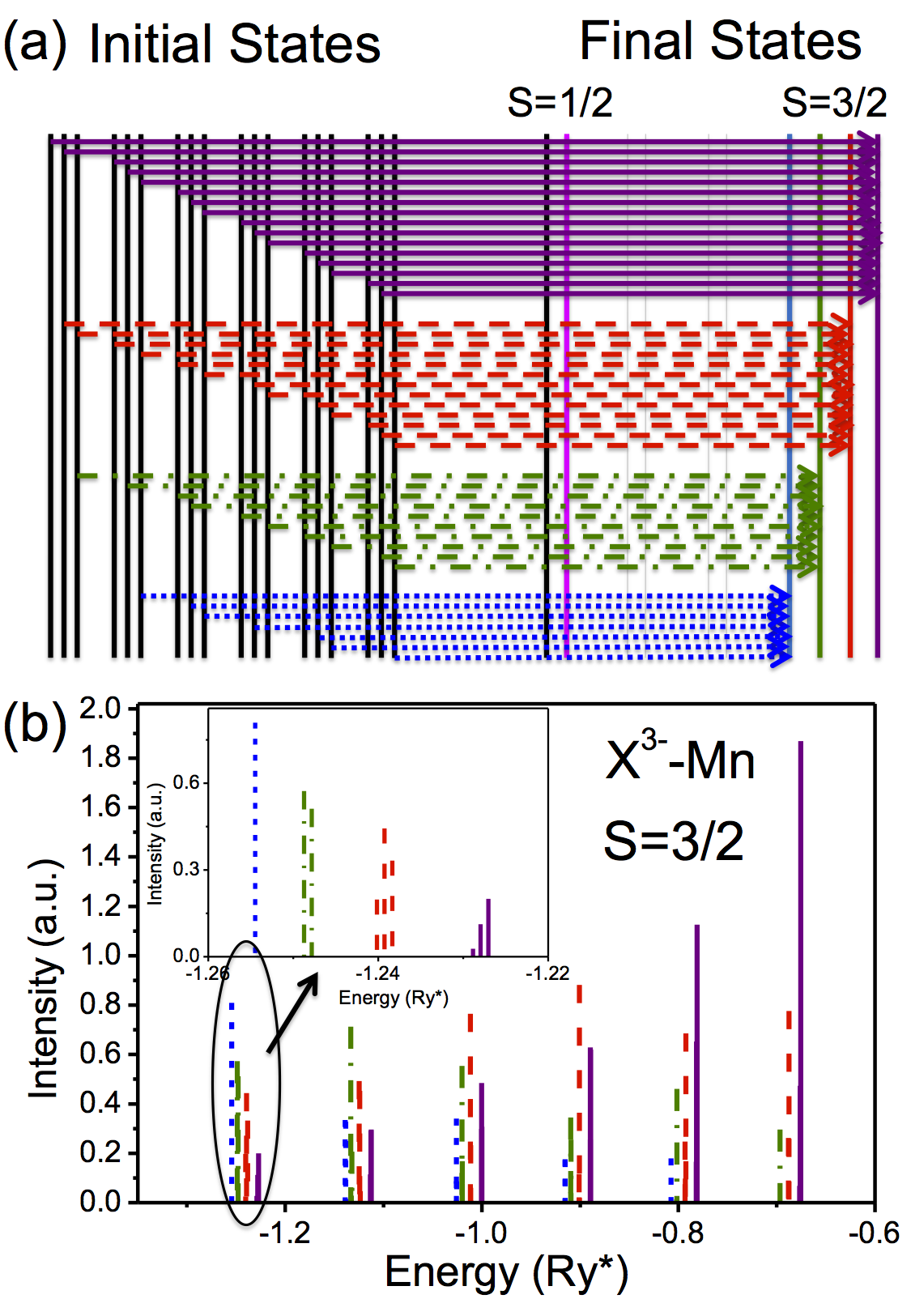}
\caption{(Color online) (a) Schematic representation of the
  $X^{3-}$-Mn initial and final states, and the allowed transitions
  from all initial states to the $S=3/2$ final states indicated by the
  arrows.  
(b) High-energy part of the $X^{3-}$-Mn emission spectrum in
$\sigma_+$ polarization. The styles of emission lines correspond to
the styles of arrows in (a). 
The inset shows the emission from the first group of three initial
states to all final states with $S=3/2$.
\label{fig7b}}
\end{center}
\end{figure}
Figure~\ref{fig7b}(a) shows the $X^{3-}$-Mn initial and final energy
levels, with the solid arrows representing optical transitions from
all initial states to four final states with $S=3/2$ (differentiated
by the color).  

In the high-energy part of the $X^{3-}$-Mn emission spectrum,
Fig.~\ref{fig7b}(b), there are also six main groups of peaks
originating from the splitting of the initial states due to the h-Mn 
interaction.  
Each of these groups is further split by the e-Mn interaction in both
the final and initial states.  
To observe these splittings, the details of the emission from the
first three energy levels of the initial state to the four final
states with $S=3/2$ are shown in the inset. 
It consists of nine emission lines (three transitions are dark), arranged
into four groups, each corresponding to a different final state.  
The splitting between these four groups corresponds to the strength of
e-Mn interaction in the final state, while the splitting within each
group is due to the indirect e-Mn interaction in the initial state.
The e-Mn-induced splitting in the final state is larger than that
in the initial state, because the final-state configuration has an
electron in an open $s$ shell directly interacting with Mn, while in
the initial state, the interaction between $p$-shell electrons and Mn
is mediated by e-e Coulomb interactions. 

\subsection{Comparison between emission spectra of different
  complexes} 

\begin{figure}
\begin{center}
\includegraphics[width=0.48\textwidth]{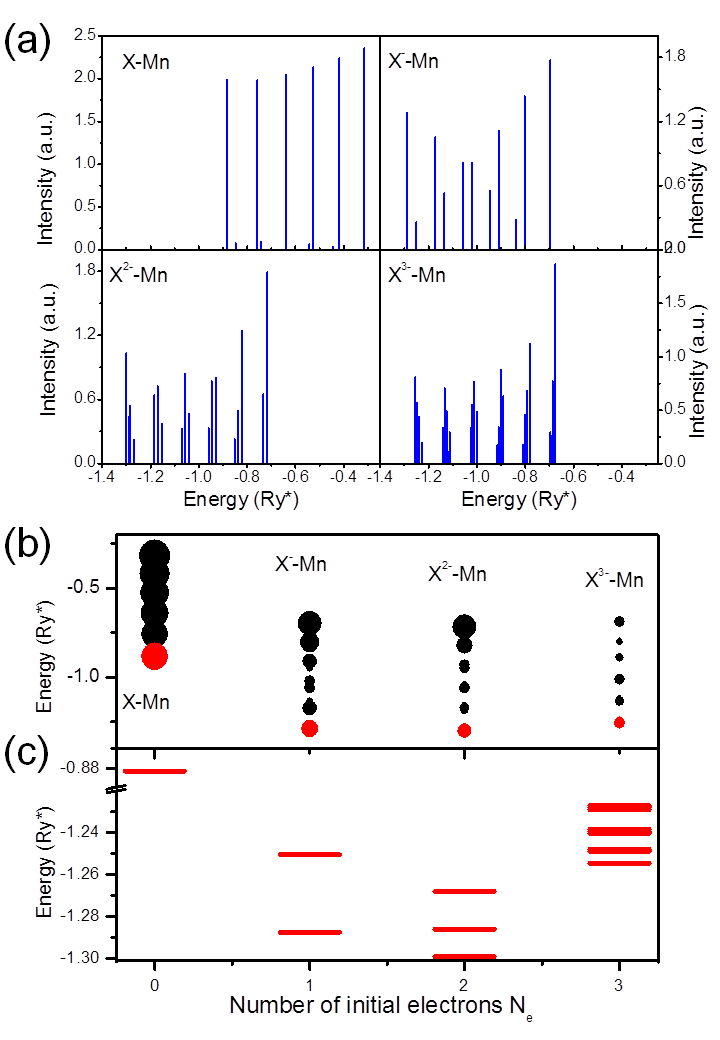}
\caption{(Color online) (a) Emission spectrum of $X$-Mn,
  $X^{-}$-Mn complexes and high-energy part of the spectrum of $X^{2-}$-Mn and $X^{3-}$-Mn complexes.
(b) Emission spectra from (a) as a function of the number of electrons in the initial state. (c) 
Close-up of the lowest-energy emission lines as a function of the number of electrons in the initial 
state with clearly visible multiplicity of lines.
\label{fig8}}
\end{center}
\end{figure}

Now we analyze the evolution of the emission spectrum for the right
circularly-polarized light as the number of excess electron $N_e$
confined in the QD is increased. 
In Fig.~\ref{fig8}(a), we show the comparison between the emission
spectra of $X$-Mn, $X^{-}$-Mn, and the high-energy parts of the emission
spectra of $X^{2-}$-Mn, and $X^{3-}$-Mn complexes. 
The emission spectra of these complexes are shown as a function of the
number $N_e$ of extra electrons in a QD  in Fig.~\ref{fig8}(b), where
the low-energy peak is clearly marked in red.  
Its energy is almost the same for $X{-}$, $X^{2-}$, and $X^{3-}$-Mn,
and it is much lower than the low-energy peak for $X$-Mn. 
This corresponds to the similar plateau as is visible in the emission
of a nonmagnetic charged
QD\cite{wojs_hawrylak_prb1997,dalacu_reimer_lpr2010}
[ see Fig. \ref{fig1}(c)].   

Figure~\ref{fig8}(c) shows the evolution of the low-energy emission
lines with the number of excess electrons $N_e$. 
For the $X$-Mn complex, there is only one emission line, since there is
only one final state of the electron-hole recombination, being the
state of Mn with $M_z=-5/2$. 
The emission from the lowest state of $X^-$-Mn complex consists of two
lines with the splitting between them corresponding to the final-state
splitting (electron-Mn complex creating state with total angular
momentum $J=2$ or $J=3$). 
In the case of the $X^{2-}$-Mn complex with $S=1$, there are three
final states of the two-electron system, all of them triplets. 
Again, the splitting between the emission lines can be translated into
the splitting between final states.  
The lowest-energy emission spectrum of the $X^{3-}$-Mn complex with
$S=3/2$ consists of four groups of levels, with the splitting
between them reflecting the splitting of the four final states of
three electrons with $S=3/2$ as described in the previous
section. 
Each of these groups exhibits a fine structure related to the fine
structure of the initial state.  

\subsection{Effects of Mn in an off-center position} 
\begin{figure}
\begin{center}
\includegraphics[width=0.48\textwidth]{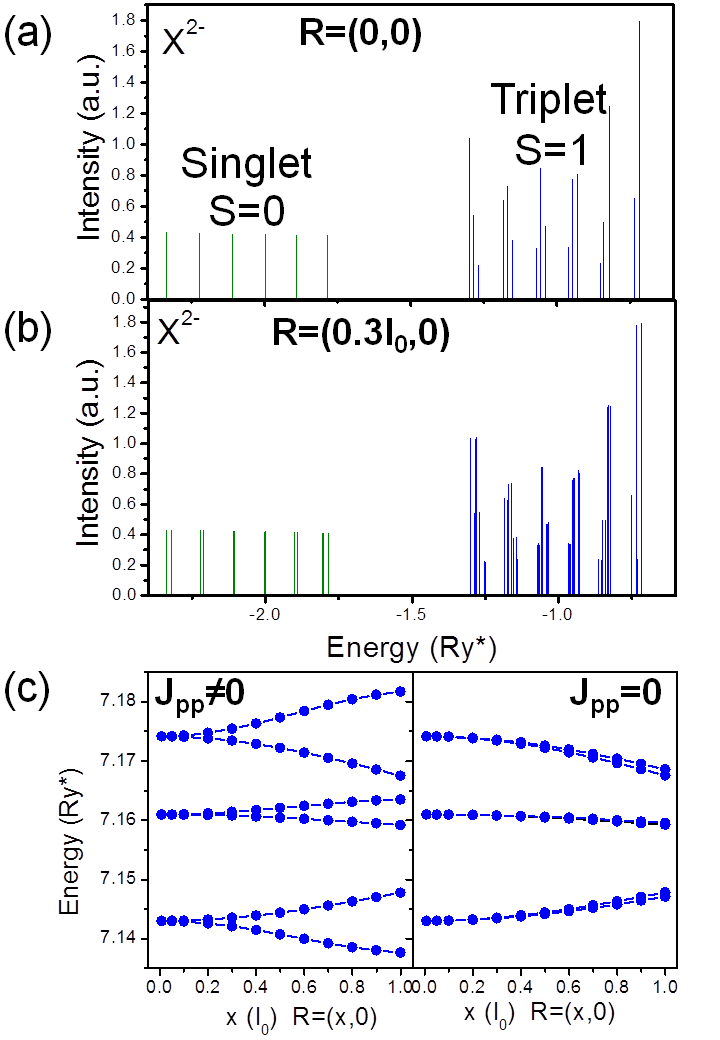}
\caption{(Color online) Emission spectrum of $X^{2-}$-Mn complex when Mn is in the center $R=(0,0)$ (a) 
or at position $R=(0.3l_{0},0)$ (b). (c) Comparison between the splitting of 2e-Mn complex in the presence 
$J_{pp}\ne 0$ (left) and absence $J_{pp}= 0$ (right) of direct
$p$-electron-Mn interaction. 
\label{fig9}}
\end{center}
\end{figure}
When Mn is positioned away from the dot center, the cylindrical
symmetry of the dot is broken and the total angular momentum $L$ is no
longer a good quantum number.
As a result, the states with finite angular momenta, e.g., $L=\pm 1$,
are coupled by Mn-induced scattering of the carriers, which 
opens additional gaps in the spectrum.
Moreover, carriers occupying QD orbitals with non-zero angular momentum,
e.g., $p$ orbitals, interact directly with Mn. 
Figures~\ref{fig9}(a) and (b) allows to compare the emission spectra
from $X^{2-}$-Mn complex confined in the QD with Mn ion in the center
$R=(0,0)$ (a) and at position  $R=(0.3 l_{0},0)$ away from the
center (b).
Indeed, in Fig.~\ref{fig9}(b), one can observe additional peaks
that arise from the removal of degeneracy of energy levels, which for a
rotationally symmetric dot were orbitally degenerate, in this case 
states with $L=\pm 1$.
Shifting Mn to a more off-center position leads to significant changes
in the emission spectrum, as discussed in
Refs.~\onlinecite{rossier_prb2006} and \onlinecite{leger_besombes_prl2005}.  
 
In previous sections, we have demonstrated that the measure of the 
strength of the indirect $p$-electron-Mn interactions is given by the 
splitting of the six groups of the emission lines in the $X^{2-}$-Mn 
spectrum, and this in turn is determined by the splitting of the final 
state of the emission. This allows to assess the relative importance 
of the indirect component of that interaction, which is the only coupling 
mechanism for the impurity in the center of the QD, compared to the 
direct $p$-electron-Mn interaction, which is present when the impurity 
is shifted off-center. In Fig.~\ref{fig9}(c), the evolution of the energies 
of the 2e-Mn complex as the Mn is displaced is presented. Figure~\ref{fig9}(c) 
(left), shows the evolution of the 2e-Mn spectrum as a function of Mn position 
capturing all direct and indirect terms, while in Figure~\ref{fig9}(c) (right) 
the direct interactions is turned off, artificially setting $J_{pp}=0$. In 
Fig.~\ref{fig9}(c) (left) the three sets of lines split as the impurity is 
shifted as due to symmetry breaking. However, apart from that, the splitting 
of 2e-Mn complex (the large gaps) is of the same order in both cases until 
the position of Mn exceeds $R\approx(0.6l_{0},0)$. We conclude that the indirect 
interaction is dominating the direct one over very broad range of the Mn position 
in the QD, and as such should not be ignored. 

\section{Summary}

In summary, we presented a microscopic theory of the optical
properties of self-assembled quantum dots doped with a single
magnetic (Mn) impurity containing a controlled number of electrons
$N_e$. 
The total spin of the electron complex is controlled by the population
of electronic shells: it is zero for closed shells and maximal for
half-filled shells. 
We show that even though electrons may occupy electronic states that are not
coupled directly with Mn, there exists an indirect coupling mediated
by electron-electron interactions.
This coupling allows for the detection of electron spin and
verification of Hund's rules in self-assembled quantum dots from
emission spectra.  
We have shown that the indirect interaction between $p$ electrons and
Mn ion is an important effect even when Mn is shifted away from the
center of the quantum  dot, and dominates over the direct interaction
over a broad range of Mn positions.
The details and a complete analysis of this e-Mn coupling mediated by
e-e Coulomb interaction is a subject of a further study. \cite{Mendes}

\section*{Acknowledgment}

The authors thank NSERC and the Canadian Institute for Advanced
Research for support. 
UCM acknowledges the support from CAPES-Brazil (Project Number
5860/11-3) and FAPESP-Brazil (Project Number 2010/11393-5).

\end{document}